\documentclass[12pt]{article}
\makeatletter
  
  \@addtoreset{equation}{section}
\makeatother

\usepackage{graphicx}

\begin{document}

\title{Oscillations and Random Perturbations of a FitzHugh-Nagumo System}
\author{Catherine Doss\thanks{
Laboratoire Jacques-Louis Lions, Bo\^ite 189, Universit{\'e} Pierre et Marie Curie-Paris 6,  4, Place Jussieu, 75252 Paris cedex 05, France; 
doss@ann.jussieu.fr }\\
\and Mich{\`e}le Thieullen\thanks{ Laboratoire de Probabilit{\'e}s et Mod{\`e}les Al{\'e}atoires, Bo\^ite 188, Universit{\'e} Pierre et Marie Curie-Paris 6,  4, Place Jussieu,
, 75252 Paris cedex 05, France; michele.thieullen@upmc.fr}\\
}

\maketitle

\begin{abstract} 
\noindent We consider a stochastic perturbation of a FitzHugh-Nagumo system. We show that it is possible to generate oscillations for values of parameters which do not allow oscillations for the deterministic system. We also study the appearance of a new equilibrium point and new bifurcation parameters due to the noisy component. 
\end{abstract}


\vfill
\eject
\section{Introduction.}

Let us consider the following family of deterministic systems indexed by the parameters $a\in{\bf R}$ and $\delta>0$. 
\begin{eqnarray}\label{FHNdet}
\delta {\dot x}_t&=&-y_t+f(x_t),\quad X_0=x\\
{\dot y}_t&=&x_t -a,\quad Y_0=y
\end{eqnarray}
and their stochastic perturbation by a one dimensional Wiener process $(W_t)$ as follows
\begin{eqnarray}\label{FHNsto}
\delta dX_t&=&(-Y_t+f(X_t))dt+\sqrt\epsilon dW_t,\quad X_0=x\\
dY_t&=&(X_t-a)dt,\quad Y_0=y
\end{eqnarray}
The function $f$ is a cubic polynomial: $f(x)= -x(x-\alpha)(x-\beta)$ with $\alpha<0<\beta$. The parameter $\delta$ is small. The deterministic system (\ref{FHNdet})-(1.2) is an example of a slow-fast system: the two variables $x,y$ have different time scales,  $ x_t$ evolves rapidly while $y_t$ evolves slowly. This system is one version of the so called FitzHugh-Nagumo system and plays an important role in neuronal modelling. In this context $x_t$ denotes the voltage or action potential of the membrane of a single neuron. It was first proposed by FitzHugh and Nagumo (cf. \cite{Fi}, \cite{N}). One interest of this model is that it reproduces periodic oscillations observed experimentally. Indeed FitzHugh-Nagumo system finds its origin in the nonlinear oscillator model proposed by van der Pol. It is also a simplification of the Hodgkin-Huxley model which describes the coupled evolution of the membrane potential and the different ionic currents: existence of different time scales enable to pass from a four dimensional model to a two dimensional one. Oscillations can take place because the deterministic system (\ref{FHNdet})-(1.2) exhibits bifurcations; more details will be given in section 3. Let us mention that oscillations in this system (\ref{FHNdet})-(1.2) can only occur when $a\in]a_0,a_1[$ where $a_0<a_1$ are two particular values of parameter $a$ namely the bifurcation parameters. \\
Our main interest in the present paper is to generate oscillations even for $a<a_0$ (symmetrically $a>a_1$) )by adding a stochastic perturbation to the deterministic system. What may be interpreted as some resonance type effect (cf. \cite{HI}, \cite{G}). We will therefore investigate possible oscillations for system (\ref{FHNsto})-(1.4). The presence of parameter $\epsilon$ introduces a third scale in the system and the relative strength of  $\delta$ and $\epsilon$  measured by  the ratio $\frac{\epsilon|\log\delta|}{\delta}$ will determine its evolution.  Our study was inspired by reference \cite{F1} where M. Freidlin considers a random perturbation of the second order equation $\delta\frac{d^2 y_t}{dt^2}=g(\frac{dy_t}{dt}, y_t)$ and performs the study of its solution using the theory of large deviations (cf. \cite{FW}). See also \cite{F2} for the study of a more general situation. In our case $g(\dot y, y)=y-f(\dot y+a)$. Although our argument is close to M. Freidlin's, the presence of parameter $a$ leads to a richer behaviour.\\
 We prove the existence of equilibrium point and limit cycles different from the deterministic ones; as in \cite{F1}, as well as a new bifurcation point  which did not exist for the deterministic system.
Our study relies on transitions between basins of attraction of stable equilibrium points due to noise. Relying on some estimation of a family of exit times(propositions 3.3 and 3.4) we study  conditions on the parameters under which a convenient stochastic dynamic system approach its main state(proposition 3.5 ),(which corresponds to the equilibrium  point exhibited in main theorem 2.2), or approach a metastable state (which corresponds to the limit cycle and to the new bifurcation parameters exhibited in main theorem 2.1).\\
 A general study of slow-fast systems perturbed by noise can be found in \cite{BG}.  Bursting oscillations in which a system alternates periodically between phases of quiescence and phases of repetitive spiking has been studied for stochastically perturbed systems in \cite{HM} and may be studied later in our stochastic setting. We recall that in the deterministic one a bursting-type behaviour has been generated  in \cite{DFP}. 
 
  \medskip
 
The paper is organized as follows. In section 2 we recall basic facts about (\ref{FHNdet})-(1.2) and we state the two main theorems (2.1) and (2.2). Section 3 is devoted to the application of large deviation theory to  (\ref{FHNsto})-(1.4) and section 4 to the proof of the main theorems.

  \section{Some Basic Results.}
  
  \subsection{Deterministic FitzHugh-Nagumo System.}
  
   \noindent In (\ref{FHNdet})-(1.2) let us consider  $\alpha<0<\beta$ and $f(x)= -x(x-\alpha)(x-\beta)$. In order to investigate the asymptotic behaviour of $(x_t,y_t)$ one first looks for equilibrium points of the system and their stability. The equilibrium points are defined as the points $(x,y)$ where the right-hand sides of both equations of the system vanish. For any value of $a$ there is therefore a unique equilibrium point for (\ref{FHNdet})-(1.2) which is $(a,f(a))$.  Moreover let $a_0, a_1,a_0< a_1$ be the two points where $f'$ vanishes. The stability of the equilibrium point $(a,f(a))$ changes when $a$ passes through the value $a_0$ (resp. $a_1$); $a_0$ and $a_1$ are called the bifurcation parameters of the system. Let us focus on $a_0$; an analogous argument holds for $a_1$.  By linearizing system (\ref{FHNdet}) at $(a_0+\eta,f(a_0+\eta))$ for $\eta$ small, we obtain system $\dot Z=AZ$ with
$$
A=\pmatrix{{\frac{f'(a_0+\eta)}{\delta}}&-\frac{1}{\delta}\cr1&0\cr}
$$  
A admits the two eigenvalues $\lambda_\pm=\frac{1}{2\delta}(f'(a_0+\eta)\pm i\sqrt{4\delta-{f'}^2(a_0+\eta)})$. The sign of $f'(a_0+\eta)$ is the same as that of $\eta$ since $f'$ is increasing in the neighbourhood of $a_0$. The point  $(a_0+\eta,f(a_0+\eta))$ is therefore an attracting (resp. repulsive) focus when $\eta<0$ (resp. $\eta>0$).  In particular, $(a,f(a))$ is stable when $a<a_0$, unstable when $a>a_0$. For $a\in]a_0,a_1[$ the system admits a limit cycle. The bifurcation is of Hopf type \cite{JPF}  It can be verified numerically that if $\delta<0.01$ the limit cycle is very close to the loop made up with the two attracting branches of the curve $y=f(x)$ where $x\mapsto f(x)$ is decreasing and $y\in[f(a_0),f(a_1)]$, and the portions of the two horizontal segments $y=f(a_0)$, $y=f(a_1)$ connecting them. When $\delta\rightarrow 0$ the period of this limit cycle is $0(1)$; by example for $f(x)=x(4-x^2)$ it is equal to $2$ (cf. \cite{Pi}).

 \subsection{Main Theorems}
 
Consider $(X_t,Y_t)$ the solution of (1.3)-(1.4) and  $S>0$ given in definition  3.3. 
Let us  assume that $\epsilon >0$ and $\delta >0$ go to zero in such a way that for some constant $c>0$.
\begin{equation}\label{hyp}
\frac{\epsilon|\log\delta|}{\delta}\rightarrow c
\end{equation}
 Let us denote by $\lim*$ any limit on $\epsilon$ and $\delta$ going to $0$ under condition (\ref{hyp}).
\newtheorem{theo}{Theorem}[section]

\begin{theo}\label{periodique}
Let  $c\in]0,S[$; then:\\
\noindent 1.If $a\in]x_{-}(c),x_{+}(c)[$ where $x_{-}(c)$ and $x_{+}(c)$ are introduced in definition 3.4; then there exist two periodic functions  $\Phi_c^a$ and $\Psi_c^a$ given in definition 4.2, s.t for all $A$, $h>0$, $y\in]f(a_0),f(a_1)[$,
\begin{eqnarray}
\lim*&{\bf P}_{(x,y)}&(\int_0^A|X_t-\Phi_c^a(t)|^2 dt>h)=0\\
\lim*&{\bf P}_{(x,y)}&(\sup_{[0,A]}|Y_t-\Psi_c^a(t)|>h)=0
\end{eqnarray}\\
\noindent 2.
 If $a<x_{-}(c)$ or $a>x_{+}(c)$ then for all $y\in ]f(a_0),f(a_1)[$, for all $h>0$ there exists ${\hat t}(y,h)$ such that for all $A>{\hat t}(y,h)$,\\
\begin{equation}
\lim*{\bf P}_{(x,y)}(\sup_{[{\hat t}(y),A]} |X_t-a|+|Y_t-f(a)|>h)=0
\end{equation}
\end{theo}

\begin{figure}[h]
\centerline{
\includegraphics[width=8cm,angle=0]{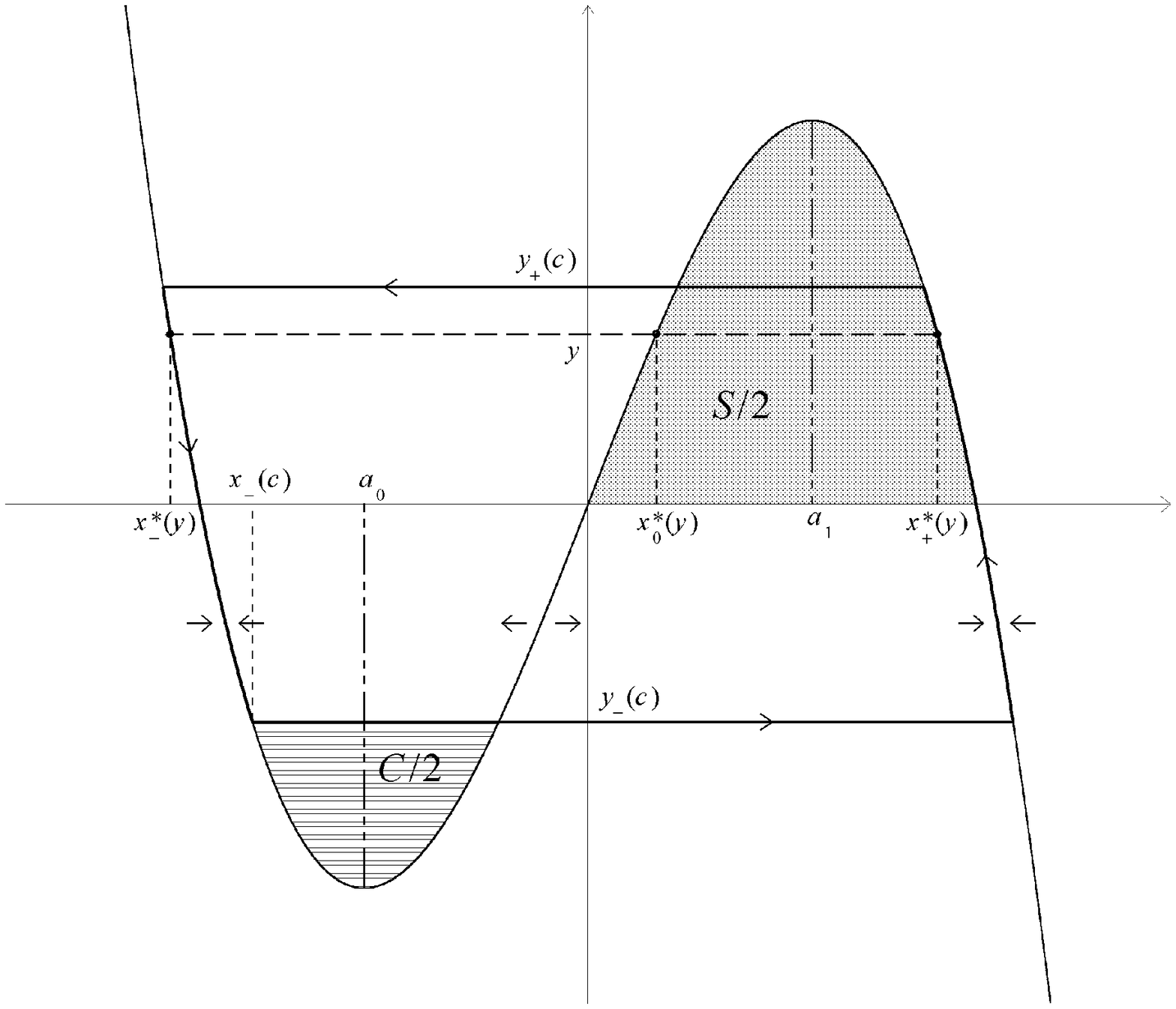}}
\caption{Limit cycle when $f(x)=x(4-x^2)$ and $\frac{\epsilon|\log\delta|}{\delta}\rightarrow c$}
\end{figure}

\noindent \textbf{Remark}\\
In the first  case the solution stabilizes when $\delta\rightarrow 0$ and $\frac{\epsilon|\log\delta|}{\delta}\rightarrow c$ around a limit cycle defined by $c$ and different from the one obtained in the deterministic case when $\delta\rightarrow 0$ and $\epsilon=0$ (see figure 1).  Moreover $ x_{-}(c)$ and by symmetry $x_{+}(c)$ play the role of bifurcation parameters for the stochastic FitzHugh-Nagumo system (\ref{FHNsto})-(1.4). Indeed for $a$ in the neighborhood of $x_{-}(c)$ but smaller  the limit of  $(X_t, Y_t)$ is a unique equilibrium point, whereas for $a$ in the neighborhood of $x_{-}(c)$ but greater it is the graph of a periodic function. These bifurcation parameters are different from those of the deterministic system (\ref{FHNdet}). This theorem is a new result w.r.t. \cite{F1}. It is made possible by the freedom on parameter $a$. The new limit cycle $(\Phi_c^a(t),\Psi_c^a(t))$ is defined in the same way as in \cite{F1} Theorem 1 Part 3., provided we take into account the presence of $a$ in our system. \\
Other regimes are considered in the work of Berglund and Gentz (cf. \cite{BG} and references therein).

\begin{theo}\label{nouvelequilibre}
Let $c> S$. Consider $x_{-}^*(y^*)$ and $x_{+}^*(y^*)$ defined in proposition 3.2 and definition 3.3; then for all $y\in ]f(a_0),f(a_1)[$ and for all $h>0$ there exists ${\hat t}(y,h)$ such that for all $A>{\hat t}(y,h)$,\\

\medskip

\noindent 1. If $a\in]x_{-}^*(y^*),x_{+}^*(y^*)[$
\begin{equation}
\lim*{\bf P}_{(x,y)}(\sup_{[{\hat t}(y),A]} |Y_t-y^*|>h)=0
\end{equation}
\noindent 2. If $a<x_{-}^*(y^*)$ or $a>x_{+}^*(y^*)$, 
\begin{equation}
\lim*{\bf P}_{(x,y)}(\sup_{[{\hat t}(y),A]}(|X_t-a|+|Y_t-f(a)|)>h)=0
\end{equation}
\end{theo}

\noindent \textbf{Remark}:\\
\noindent Case 1 may be considered as a degenerate version of the limit cycle of case 1 theorem 2.1. In fact $y^*$ is a fixed point but $X_t$ oscillates between $x_{-}^*(y^*)$ and $x_{+}^*(y^*)$.

\section{Exit Time, Main State and Metastable State}
  \subsection{Basic results on Large Deviations Theory} 
 Because of the slow-fast property of FitzHugh-Nagumo systems, the slow variable $Y_t$ of system (1.3)-(1.4) may be in a first approximation frozen at the value $y$ which leads us to the study of the family of one dimensional dynamical systems indexed by parameter $y$,which plays a basic role in the study of FitzHugh-Nagumo systems (\ref{FHNdet})-(1.2) and (\ref{FHNsto})-(1.4):
  \begin{equation}\label{detyfixe}
  dx_t^y=(-y+f(x_t^y))dt,\quad x_0^y=x
  \end{equation}
   
 \noindent So we are led to consider the real valued deterministic system 
  \begin{equation}\label{sysdet}
  dx_t=b(x_t)dt,\quad x_0=x
  \end{equation}
  and its perturbation by a brownian motion
  \begin{equation}
  d{\tilde x}_t=b({\tilde x}_t)dt+\sqrt{\tilde\epsilon} dW_t, \quad {\tilde x}_0=x
  \end{equation}
  
 \noindent  Let us briefly recall some results from \cite{FW}.
The process $({\tilde x}_t)$ describes the movement of a particle on the real line submitted to the force field $b(x)$ and to a stationnary Gaussian noise of amplitude $\sqrt{\tilde\epsilon}$. When $\tilde\epsilon\rightarrow 0$, $({\tilde x}_t)$ converges to the solution $(x_t)$ of (\ref{sysdet}): 
  \begin{equation}
  \forall\eta>0\quad\forall T>0\quad \lim_{{\tilde\epsilon}\rightarrow 0}{\bf P}(\sup_{[0,T]}|{\tilde x}_t -x_t|>\eta)=0
  \end{equation}  
  However because of diffusion due to the presence of noise, some trajectories of the process $({\tilde x}_t)$ may present large deviations from those of the deterministic system $(x_t)$. Such deviations are measured by means of the action functional $S_{T_1}^{T_2}(\varphi)$ independant of $\tilde\epsilon$ and defined by
  \begin{equation}
 S_{T_1}^{T_2}(\varphi)=\frac{1}{2}\int_{T_1}^{T_2} | {\dot\varphi}_u-b(\varphi_u)|^2 du
  \end{equation}
 when $\varphi$ is absolutely continuous, by $S_{T_1}^{T_2}(\varphi)=+\infty$ otherwise
 
 \begin{theo}
  ( \cite{FW}, Lemma 2.1, Chap. 4)\\
   Let $\eta>0$. Then
\begin{equation}
{\bf P}_x(\sup_{[0,T]}|{\tilde x}_t-x_t|\geq\eta)\leq\exp(-\frac{1}{\tilde\epsilon}[\inf_{\Delta} S_0^T(\varphi)+o(1)])
\end{equation}
when $\tilde\epsilon\rightarrow0$ and where $\Delta:=\{\varphi; \varphi_0=x,   \sup_{[0,T]}|\varphi_t-x_t|\geq\eta\}$. 
\end{theo}
Large deviations theory also provides estimates on the first exit time of $({\tilde x}_t)$ from a domain (cf.  \cite{FW}, Theorem 4.2, Chap. 4). Domains of interest are basins of attraction of stable equilibrium points of (\ref{sysdet}). The key tools are quasipotentials. 
 \newtheorem{definition}{Definition}[section]
\begin{definition}
The quasi potential of the deterministic system (\ref{sysdet}) w.r.t. a point $\overline x$ (also called transition rate) is defined as the function 
\begin{equation}
u\mapsto V(u):=\inf \{S_{T_1}^{T_2}(\varphi);0\leq T_1<T_2, \varphi(T_1)=\overline x, \varphi(T_2)=u\}
\end{equation}
\end{definition}

 \newtheorem{proposition}{Proposition}[section]
 
 \begin{proposition}\label{onedim}
  The quasi potential of (\ref{sysdet}) w.r.t. $\overline x$ coincides with the function 
  \begin{equation}
  u\mapsto V(u)=-2\int_{\overline x}^u b(r)dr
  \end{equation}
 \end{proposition}

 \noindent  \textbf{Remark}:\\
 \noindent The above statement holds since (\ref{sysdet}) is one dimensional. It also holds in the multidimensional case when the drift $b$ of (\ref{sysdet}) is a gradient.

\begin{theo}\label{estimeesortie}
Let $x^*$ be a stable equilibrium point of (\ref{sysdet}) such that $b(r)<0$ for all $r>x^*$, and $b(r)>0$ for all $ r<x^*$. Let $D$ be the basin of attraction of $x^*$ and $\tilde\tau$ denote the first exit time of ${\tilde x}$ from $D$. Let us assume that $D=]\alpha_1,\alpha_2[$ with $V(\alpha_1)<V(\alpha_2)$. Then for all $ x\in D$ and $ h>0$,
\begin{equation}
\lim_{\tilde\epsilon\rightarrow 0}{\bf P}_x({\tilde x}_{\tilde\tau}=\alpha_1)=1
\end{equation}
\begin{equation}\label{tempssortie}
\lim_{\tilde\epsilon\rightarrow 0}{\bf P}_x({\rm e}^{\frac{V(\alpha_1)-h}{\tilde\epsilon}}<\tilde\tau<{\rm e}^{\frac{V(\alpha_1)+h}{\tilde\epsilon}})=1
\end{equation} 
 \end{theo}

\noindent \textbf{Remark}:\\
\noindent With the notations of Theorem \ref{estimeesortie}, with great probability when $\tilde\epsilon\rightarrow 0$, the behaviour of the process $({\tilde x}_t)$ on an interval $[0,T(\tilde\epsilon)]$, in particular whether the process has jumped out of the basin of attraction $D$ or not, depends on the value of $\tilde\epsilon\log T(\tilde\epsilon)$ compared to $V(\alpha_1)$.

 \subsection{Two Families of Quasipotentials.}
 
Let us now apply these results to the following family of one dimensional dynamical systems indexed by parameter $y$ introduced in   the preceeding subsection.
\begin{equation}\label{detyfixe}
  dx_t^y=(-y+f(x_t^y))dt,\quad x_0^y=x
  \end{equation}

 \begin{proposition}\label{equilibriumpoints}
  Let  $y \in]f(a_0),f(a_1)[ $ with $a_0$ and $a_1$  the two points where $f'$ vanishes.  
  
  \noindent (i) The set $\{x\in{\bf R}; f(x)=y\}$ consists of three points $x_{-}^*(y)<x_0^*(y)<x_{+}^*(y)$ (the equilibrium points of (\ref{detyfixe})) each being a continuous function of $y$ with bounded first and second derivatives. 
  
  \noindent (ii) The two points  $x_{\pm}^*(y)$ are stable. $x_0^*(y)$ is unstable. 
 \end{proposition}
 
\noindent \textbf{Proof of Proposition \ref{equilibriumpoints}}:\\
 Left to the reader; we refer to figure 1-section 2
 
 \begin{definition} Let us define the two functions $V_{\pm}$ on $]f(a_0),f(a_1)[$ as follows:
 \begin{equation}\label{quasipotentiels}
 V_{\pm}(y)=-2\int_{x_{\pm}^*(y)}^{x_0^*(y)}(-y+f(u))du.
 \end{equation}
 \end{definition}
  \noindent From Proposition \ref{onedim} we see that $V_{\pm}(y)$ is the value at $x_0^*(y)$ of the quasipotential of (\ref{sysdet}) w.r.t. $x_{\pm}^*(y)$. Both functions  $ V_{\pm}(y)$ are strictly monotone: $ V_{-}$ is strictly increasing,  $ V_{+}$ is strictly decreasing. Therefore their graphs restricted to $]f(a_0),f(a_1)[$ intersect at a unique point.

 \begin{definition}\label{notations1}
\noindent  We denote by $(y^*,S)$ the intersection point of the graphs of $ V_{-}$ and $ V_{+}$. Let ${\cal E}_1:=\{y>y^*\}$ and ${\cal E}_2:=\{y<y^*\}$.
\end{definition}

\begin{definition}\label{notations2}
\noindent For $c\in]0,S[$ we denote by $y_{\pm}(c)$  the points of $]f(a_0),f(a_1)[$ which satisfy $y_{-}(c)< y^*<y_{+}(c)$ and $ V_{-}(y_{-}(c))=c=V_{+}(y_{+}(c))$. Let us also  define $x_{-}(c):=x_{-}^*(y_{-}(c))$ and  $x_{+}(c):=x_{+}^*(y_{+}(c))$ (cf. figure 1).
\end{definition}

\noindent\textbf{Remark}:\\
\noindent By definition $ V_{-}(y^*)=V_{+}(y^*)=S$. For  $f(x)=4x-x^3$, $y^*=0$ and $S=4$. 

 \medskip

 \noindent For any function $U$ satisfying $y-f(u)\equiv{-\partial_x U}/2 $ the following identities hold:
\begin{equation}\label{wells}
V_{-}(y)=U(x_0^*(y))-U(x_{-}^*(y)), \quad V_{+}(y)=U(x_0^*(y))-U(x_{+}^*(y))
\end{equation}

\noindent In our case such a function $U$ is a polynomial of degree $4$ which admits $x_0^*(y)$ as relative maximum and $x_{\pm}^*(y)$ as relative minima. The graph of the function $U$ has two wells with respective bottoms at $x_{\pm}^*(y)$ and one top at $x_0^*(y)$. Identities (\ref{wells}) express that $V_{\pm}(y)$ are the respective depths of these wells. Therefore, on ${\cal E}_1$ the well with bottom $x_{-}^*(y)$ is the deepest one while it is the contrary on ${\cal E}_2$.

\noindent Second, the portion of the curve $z=f(x)$ connecting $(x_{-}^*(y),y)$ to $(x_0^*(y),y)$ is situated below the horizontal line $L_y:=\{(x,z);z=y\}$; thus the positive quantity $\frac{1}{2}V_{-}(y)$ measures the surface of the area limited by the curve $z=f(x)$ and the segment of $L_y$ on which $x\in [x_{-}^*(y) ,x_0^*(y)]$. In the same way $\frac{1}{2}V_{+}(y)$ measures the surface of the area limited by the curve $z=f(x)$ and the segment of $L_y$ on which $x\in [x_0^*(y),x_{+}^*(y) ]$ but in this case the portion of the curve is above the line segment. 

\noindent Moreover as we will see in the following section $V_{\pm}$ is connected to exit times of diffusions
 \begin{equation}
d{\tilde Z}_t^y=(-y+f({\tilde Z}_t^y))dt+\sqrt{\tilde\epsilon} dW_t,\quad {\tilde Z}_0^y=x
\end{equation}
from the basins of attraction of $x_{\pm}^*(y)$ (cf.Theorem \ref{estimeesortie}).

\subsection{Exit Times, Main state and Metastable States}

We refer the reader to \cite{F2} for the present section. Let us recall the fundamental difference between the two parameters $\epsilon$ and $\delta$. Parameter $\delta$ is already present in the deterministic system (\ref{FHNdet})-(1.2) where it measures the difference between the time scale of the slow variable $y_t$ and the time scale of the fast variable $x_t$. In particular after the time change $s:=\delta t$, the trajectory $({\tilde x}_s,{\tilde y}_s):=(x_{\delta t},y_{\delta t})$ satisfies
\begin{eqnarray}\label{dettime}
{\dot {\tilde x}}_s&=&-{\tilde y}_s+f({\tilde x}_s)\\
{\dot {\tilde y}}_s&=&\delta({\tilde x}_s -a)
\end{eqnarray}
Since $\delta$ is small  the component ${\tilde y}_s$ may be considered as constant equal to ${\tilde y}_0$. Let us define $Z_t^y$ to be the family of solutions of 
\begin{equation}
\delta d Z_t^y=(-y+f( Z_t^y))dt+\sqrt\epsilon dW_t,\quad  Z_0^y=x.
\end{equation}
We will also consider the time change $({\tilde Z}_t^y)$ of $(Z_t^y)$ under $s:=\delta t$
\begin{equation}\label{Ztilda}
d{\tilde Z}_s^y=(-y+f({\tilde Z}_s^y))ds+\sqrt{\tilde\epsilon} d{\tilde W}_s,\quad {\tilde Z}_0^y=x
\end{equation}
with $\tilde\epsilon=\frac{\epsilon}{\delta}$. $({\tilde Z}_t^y)$ is the stochastic perturbation of (\ref{detyfixe}). \\
In the two following propositions we give crucial estimates on exit  time of  the respective solutions of (3.17) and (3.18). 

\begin{proposition}\label{tempssortie1}
Let ${\tilde\tau}_1^y$ (resp. ${\tilde\tau}_2^y$) denote the exit time of ${\tilde Z}^y$ from $D_1^y$ (resp. $D_2^y$) which is the basin of attraction of $x_{-}^*(y)$ (resp. $x_{+}^*(y)$). Let us recall that $D_1^y =]-\infty,x_0^*(y)[$ (resp. $D_2^y=]x_0^*(y),+\infty[$).  From Theorem \ref{estimeesortie} identity (\ref{tempssortie}), for $x\in D_1^y$ and $h>0$ we obtain\\
\noindent $\forall x \in ]-\infty,x_0^*(y)[ , \forall h>0$ 
\begin{equation}\label{tildeexit1}
\lim_{\tilde\epsilon\rightarrow 0}{\bf P}_x(\exp(\frac{V_{-}(y)-h}{\tilde\epsilon})<{\tilde\tau}_1^y<\exp(\frac{V_{-}(y))+h}{\tilde\epsilon}))=1
\end{equation} 
An analogous result holds for $x\in D_2^y$ by replacing  $\tilde\tau_1^y$ by $\tilde\tau_2^y$ and $V_{-}(y)$ by $V_{+}(y)$:\\
\noindent $\forall x \in ]x_0^*(y),+\infty[ , \forall h>0 $:
\begin{equation}\label{tildeexit2}
\lim_{\tilde\epsilon\rightarrow 0}{\bf P}_x(\exp(\frac{V_{+}(y)-h}{\tilde\epsilon})<{\tilde\tau}_2^y<\exp(\frac{V_{+}(y))+h}{\tilde\epsilon}))=1
\end{equation} 

\end{proposition}

\begin{proposition}\label{tempssortie2}
Let $\tau_1^y$ (resp. $\tau_2^y$) denote the first exit time of $Z^y$ from $D_1^y$ (resp. $D_2^y$). The law of $\tau_i^y$ is the same as the law of $\delta{\tilde\tau}_i^y$ for $i=1,2$. Let us assume that $\epsilon$ and $\delta$ go to $0$ in such a way that 
\begin{equation}\label{epsilondelta}
\frac{\epsilon}{\delta}|\log\delta|\rightarrow c\in]0,+\infty[.
\end{equation}
In this case we obtain: \\ 
\noindent $\forall x \in ]-\infty,x_0^*(y)[ , \forall h>0$
\begin{equation}\label{sortiebassin1}
\quad\lim{\bf P}_x(\delta^{c^{-1}(c-V_{-}(y)+h)}<\tau_1^y<\delta^{c^{-1}(c-V_{-}(y)-h)})=1
\end{equation} 
An analogous result holds for $x\in D_2^y$ by replacing  $\tau_1^y$ by $\tau_2^y$ and $V_{-}(y)$ by $V_{+}(y)$:\\
\noindent$ \forall x \in  ]x_0^*(y),+\infty[, \forall h>0$
\begin{equation}\label{sortiebassin2}
\quad\lim{\bf P}_x(\delta^{c^{-1}(c-V_{+}(y)+h)}<\tau_2^y<\delta^{c^{-1}(c-V_{+}(y)-h)})=1
\end{equation} 

\end{proposition}

These two properties enable us to introduce some remarks about the main theorems stated in the previous section. These remarks are linked to the notions of mainstate and metastable state introduced in \cite{FW}.\\
In a general framework ( cf. \cite{F2}) the main state is the point towards which the cost of moving, or the transition rate, is minimum. It is not always unique: for instance in our bistable case there are two main states when $y=y^*$. Before reaching the main state, the process may reach metastable ones accessible for shorter time lenghts. \\
Main states may be considered as stable states, whereas metastable states are only stable in some time scale.

To study transitions of (\ref{Ztilda}) between the two basins of attraction during $[0,T(\tilde\epsilon)]$, the relevant quantity to consider is $\tilde\epsilon\log T(\tilde\epsilon)$ which we must compare to the transition rates $V_{\pm}(y)$ given by (\ref{quasipotentiels}). 
\noindent For the time scale $T(\tilde\epsilon)={\rm e}^{\frac{c}{\tilde\epsilon}}$; $\tilde\epsilon\log T(\tilde\epsilon)=c$. So we must compare $c$ to the transition rates $V_{\pm}(y)$. Actually this amounts to compare first $c$ to $S$ defined in definition 3.3. We refer again to figure 1.\\

More precisely we can state using definition 3.3 and 3.4:

\begin{proposition}\label{state}; 
\noindent 1. When $c>S$ the main state of ${\tilde Z}^y$ is equal to $x_{+}^*(y)$ (resp. $x_{-}^*(y)$) if $y<y^*$ (resp. $y>y^*$). \\
When $y=y^*$ the two points $x_{\pm}^*(y^*)$ are both main state.\\
\noindent 2. When $c<S$ for $y \in ]y_{-}(c),y_{+}(c)[$ and $x\in D_1^y$ (resp. $x\in D_2^y$) the metastable state of ${\tilde Z}^y$ is equal to  $x_{-}^*(y)$ (resp.  $x_{+}^*(y)$).
\end{proposition}

\noindent \textbf{Proof of Proposition \ref{state}}:\\
\noindent Direct consequence of the estimation of the time of exit given in proposition 3.4.

\noindent \textbf{Remark}:\\
\noindent 1. When $c>S$ the time interval $[0,T(\tilde\epsilon)]$ is long enough so that the process $Z^y$ reaches with great probability a small neighborhood of its main states.\\
 And as we can find an open interval $I$ containing  $y^*$ such that:
\begin{equation}\label{lemme1}
\forall y\in I\quad c>\max(V_{-}(y),V_{+}(y)). 
\end{equation}
 following proposition \label{tempssortie 2} both exit time $\tau_1^y$ and $\tau_2^y$ tends to 0 in probability so we can only expect results on the slow component and the result depends on whether the boundary between the two main states is attracting or notas is shown in theorem(2.2).\\
 \noindent 2. On the countrary when $c<S$ the time interval $[0,T(\tilde\epsilon)]$ is too short and the process only reaches with great probability a neighborhood of a metastable state.
 In this case we have $y_{-}(c)<y_{+}(c)$ and if we  consider the interval $]y_{-}(c),y_{+}(c)[$ when $c<V_{-}(y)<V_{+}(y)$ and $x\in D_1^y$ (resp.  $c<V_{-}(y)<V+{-}(y)$ and $x\in D_2^y$), for all $y\in ]y_{-}(c),y_{+}(c)[$ exit time $\tau_1^y$ (resp. $\tau_2^y$ )tends to infinity. So the process $Z^y$ remain in a neighborhood of one metastable state and switch to the other one as soon as $Y_t$ gets out of $]y_{-}(c),y_{+}(c)[$; this give rise to a limit cycle as is shown in theorem (2.1).

\section{Proof of the Main Theorems}

\begin{definition}\label{period}
For $a\in ]x_{-}(c),x_{+}(c)[$ and $y\in ]y_{-}(c),y_{+}(c)[$ as we can check on figure 1 we have:\\
\begin{equation}
x_{-}^*(z)<x_{-}(c)<a<x_{+}(c)<x_{+}^*(z)
\end{equation}
Then for $c\in]0,S[$, $a\in]x_{-}(c),x_{+}(c)[$ and $y\in ]y_{-}(c),y_{+}(c)[$, the following definitions of time duration makes sense
\begin{equation}
T_1^a(c)=\int_{y_{-}(c)}^{y_{+}(c)}\frac{dy}{x_{+}^*(y)-a}\quad {\rm and}\quad T_2^a(c)=\int_{y_{-}(c)}^{y_{+}(c)}\frac{dy}{|x_{-}^*(y)-a|}
\end{equation}
where $y_{\pm}(c)$ and $x_{\pm}(c)$ have been introduced in Proposition 3.2 and Definition \ref{notations2}.
\end{definition}

On the interval $[0,T_1^a +T_2^a]$we can introduce the solution of the following ordinary differential equations used to define the limit cycle in Theorem\label{periodique}
\begin{definition}\label{notationsder}
For $c\in]0,S[$, $a\in]x_{-}(c),x_{+}(c)[$ and $y\in ]y_{-}(c),y_{+}(c)[$ we define  $\Psi_c^a$ as the continuous periodic function with period $T^a(c)=T_1^a(c)+T_2^a(c)$ satisfying $\Psi_c^a(0)=y_{-}(c)$ and the ode
\begin{eqnarray}\label{odes}
{\dot\Psi}_c^a(t)&=&x_{+}^*(\Psi_c^a(t))-a , \quad t\in[0,T_1^a(c)[\\
{\dot\Psi}_c^a(t)&=&x_{-}^*(\Psi_c^a(t))-a , \quad t\in[T_1^a(c),T_1^a(c)+T_2^a(c)[
\end{eqnarray}
For $t\notin \{ kT^a(c), kT^a(c)+T_1^a(c), k\in{\bf Z}\}$, we denote by $\Phi_c^a(t)$ the derivative of $\Psi_c^a$ at $t$. 
\end{definition}
Existence and regularity of solutions to odes (4.3)-(4.4) follow from the $C^2$ regularity of the functions $x_{\pm}.$. Under the assumptions we recall that we have (see figure 1)
\begin{equation}
x_{-}^*(z)<x_{-}(c)<a<x_{+}(c)<x_{+}^*(z)
\end{equation}

so $\Psi_c^a(t)$ increases from $y_{-}(c)$ to $y_{+}(c)$ in a duration of time equal to $T_1^a(c)$ and decreases from $y_{+}(c)$ to $y_{-}(c)$ in a duration of time equal to $T_2^a(c)$. The periodic function $\Psi_c^a$ is obtained by continuously sticking together at $y_{\pm}(c)$ the solutions ${\overline y}_{\pm}^a$ of the following odes
\begin{equation}
{\dot{\overline y}}_{\pm}^a(t)=x_{\pm}^*({\overline y}_{\pm}^a(t))-a,\quad {\overline y}_{\pm}^a(0)=y
\end{equation}

\noindent \textbf{Remark}:\\
\noindent The function  $\Psi_c^a$ is continuous on ${\bf R}$ differentiable with continuous derivatives except at points in the set $\{kT^a(c), kT^a(c)+T_1^a(c), k\in{\bf Z}\}$.\\
 For $ t\in[0,T_1^a(c)[$ the point $(x_{+}^*(\Psi_c^a(t)), \Psi_c^a(t))$ belongs to the right stable branch;  for $t\in[T_1^a(c), T_1^a(c)+T_2^a(c)[ $ the point $(x_{-}^*(\Psi_c^a(t)), \Psi_c^a(t))$ belongs to the left stable branch.

\bigskip

\noindent{\bf Proof of Theorem \ref{periodique}.} \\
\noindent 1. Case $a\in]x_{-}(c),x_{+}(c)[$:\\
a) Let $y<y_{-}(c)$ and  $x\in D_1^y$. Then $\lim*{\bf P}_x(\tau_1^y=0)=1$ since $V_{-}(y)<V_{-}(y_{-}(c))=c$ (remember that  $V_{-}$ is strictly increasing). Therefore the process $(X_t)$ leaves $D_1^y$ instantaneously and is attracted to a neighborhood of $x_{+}^*(y)>x_{+}(c)$.  The point $(X_t, Y_t)$ remains in a neighbourhood of the branch $v=f(u)$ containing $(x_{+}^*(y),y)$ and $Y_t$  increases as long as $X_t>x^+(c)$ since $dY_t=(X_t-a)dt$. However identity (3.23) implies that $\lim*{\bf P}_x(\tau_2^z=+\infty)=1$ for $z<y_{+}(c)$ (resp. $\lim*{\bf P}_x(\tau_2^z=0)=1$ for $z>y_{+}(c)$ ). Therefore the point $(X_t,Y_t)$ is instantaneously attracted to a neighborhood of $(x_{-}^*(y_{+}(c)),y_{+}(c))$ after $Y_t$ has crossed $y_{+}(c)$ since the speed of $Y_t$ is strictly positive in a neighbourhood of $(x_{+}(c),y_{+}(c))$. The argument is then the same as before. We detail it for completeness. Since  $x_{-}^*(y_{+}(c))<x_{-}(c)<a$, the second coordinate $Y_t$ decreases as long as $X_t-a<0$. However identity (3.22) implies that $\lim*{\bf P}_x(\tau_1^z=+\infty)=1$  for $z>y_{-}(c)$ (resp. $\lim*{\bf P}_x(\tau_1^z=0)=1$ for $z<y_{-}(c)$). Therefore the point $(X_t,Y_t)$ is instantaneously captured by $(x_{+}^*(y_{-}(c)),y_{-}(c))$ after $Y_t$ has crossed $y_{-}(c)$. Hence $(X, Y)$ converges in probability to a limit cycle of period $T^a(c)=T_1^a(c)+T_2^a(c)$.\\
b) Let $y>y_{-}(c)$  and  $x\in D_1^y$. By identity 3.22  $\lim*{\bf P}_x(\tau_1^y=+\infty)=1$, the fast process $(X_t)$ is attracted to $x_{-}^*(y)$. After this first phase, one can apply the same argument as in a).\\

\noindent 2.Case $a<x_{-}(c)$. As already mentioned in section 3, when $c<S$, the relevant states are the metastable ones. The assumption $a<x_{-}(c)$ implies $f(a)>y_{-}(c)$. Let ${\cal V}$ be a small neighborhood ${\cal V}$ of $(a,f(a))$. Let us first assume that ${\cal V}$ is so small that all $(u,v)\in {\cal V}$ satisfies $v>y_{-}(c)$ and accordingly $u<x_{-}(c)$. Let $(X_t,Y_t)$ start from $(u,v)$. In particular $u\in D_1^v$ and $\tau_1^v=+\infty$ since $V_{-}(v)>c$. Then $(Y_t)$ will evolve like the solution of ${\dot v}_t=x_{-}^*(v_t)-a$, $v_0=v$ for which $f(a)$ is a stable equilibrium point. Let us now assume that there are points $(u,v)\in {\cal V}$ such that $v<y_{-}(c)$.  Then $(X_t)$ instantaneously jumps to the right stable branch; $(Y_t)$ becomes close to the solution of ${\dot v}_t=x_{+}^*(v_t)-a$ and therefore increases until it reaches $y_{+}(c)$. At that time it jumps to the left stable branch and we are back to the previous argument since then $(Y_t)$ becomes close to the solution of ${\dot v}_t=x_{-}^*(v_t)-a$.\\
 \noindent Let us notice that for the slow variable $Y_t$ we get a result using the uniform norm by large deviation estimates but  for the quick variable $X_t$ the  result is formulated in $L^1$ norm thanks to  equation (1.4).
 \bigskip

\noindent {\bf Proof of Theorem \ref{nouvelequilibre}.} When $c>S$, as pointed out in subsection (3) $Z^y$ has time to reach its main state so the evolution of $Y_t$ is close to the solution of
\begin{eqnarray*}
{\dot v}_t&=&x_{+}^*(v_t)-a\quad {\rm provided}\quad  v_t<y^*\\
{\dot v}_t&=&x_{-}^*(v_t)-a\quad {\rm provided} \quad v_t>y^*
\end{eqnarray*}
and it depends on whether the boundary $\{ y^*\}$ between ${\cal E}_1:=\{y>y^*\}$ and ${\cal E}_2:=\{y<y^*\}$ is attractive for this system or not(cf. \cite{F2}). If $a\in ]x_{-}^*(y^*),x_{+}^*(y^*)[$ the boundary is attractive. This is not the case when $a<x_{+}^*(y^*$ nor when $a>x_{+}^*(y^*)$. 

\medskip

\noindent 1. Let $x_{-}^*(y)<a<x_{+}^*(y^*)$. The point $y^*$ is an attracting boundary since $x_{-}^*(y^*)-a<0<x_{+}^*(y^*)-a$. 

\noindent Assume for example that $Y_0=y<y^*$ that is $y\in{\cal E}_2$. The evolution of $(Y_t)$ is close to the solution of ${\dot{\overline Y}}_t=x_{+}^*({\overline Y}_t)-a$. Therefore since $x_{+}^*(y)>a$, process $(Y_t)$ starts increasing until it reaches $y^*$. After it has reached $y^*$, the evolution of $(Y_t)$ is close to the solution of ${\overline Y}_0=y^*$, ${\dot{\overline Y}}_t=B({\overline Y}_t)$ where $B$ is a vector field tangent to boundary (cf. \cite{F2} Theorem 1)  which is the $0$-dimensional manifold $\{y^*\}$; therefore $B$ is zero and ${\overline Y}_t\equiv y^*$. The time ${\hat t}(y,h)$ is the time necessary to reach a small ball $B(y^*,h)$.

\noindent If $y\in{\cal E}_1$ the evolution of $(Y_t)$ is close to the solution of ${\dot{\overline {\cal Y}}}_t=x_{-}^*({\overline {\cal Y}}_t)-a$. In this case process $(Y_t)$ starts decreasing since $x_{-}^*(y)<a$ until it reaches $y^*$. The argument and the conclusion are then identical to the preceeding case. 

\medskip

\noindent 2. Let $a>x_{+}^*(y)$. The point $y^*$ is not an attracting boundary since $x_{-}^*(y^*)-a=<0$ and $x_{+}^*(y^*)-a<0$. 

\noindent Let us assume for example that $Y_0=y\in {\cal E}_2$. The evolution of $(Y_t)$ is close to the solution of ${\dot{\overline Y}}_t=x_{+}^*({\overline Y}_t)-a$. The value $f(a)$ is a stable equilibrium for $({\overline Y}_t)$. If $x_{+}^*(y)>a$, process $(Y_t)$ starts increasing until it reaches $f(a)$ to which it is attracted. If now  $x_{+}^*(y)<a$ the argument and the conclusion are the same except that $(Y_t)$ starts decreasing. If $y\in {\cal E}_1$ the evolution of $(Y_t)$ is close to the solution of ${\dot{\overline {\cal Y}}}_t=x_{-}^*({\overline {\cal Y}}_t)-a$. Since $x_{-}^*(y^*)-a<0$ process$(Y_t)$ starts decreasing until it crosses $y^*$ (which is not attractive) towards ${\cal E}_2$. Then we are back to the previous case. 

\medskip

\noindent 3. The case $a<x_{-}^*(y)$ is treated analogously.

 \bigskip
 
 \noindent {\bf Acknowledgment.} We thank Prof. Mark Freidlin for sending us a copy of his paper \cite{F2}  which was not available to us.


\begin{thebibliography}{99}

\bibitem{BG} Berglund, N., Gentz, B., {\it Noise-Induced Phenomena in Slow-Fast Dynamical Systems: A Sample Paths Approach.} Springer (2006)

\bibitem{DFP} Doss, C., Fran{\c c}oise, J.-P., Piquet, C., {\ it Bursting oscillations in two coupled Fitzhugh-Nagumo system.} ComplexUs, vol.2, p 1-11 (2003)

\bibitem{Fi} FitzHugh, R., {\it Impulses and physiological states in theoretical models of nerve membrane.} Biophys. J., 445-466 (1961)

\bibitem {JPF} Fran{\c c}oise, J.-P. {\it Oscillations en biologie. Analyse qualitative et mod\`eles.} Collection Math\'ematiques et Applications Vol. 46, SMAI Springer 2005


\bibitem {FW} Freidlin, M.,I.,  Wentzell, A., D., {\it Random perturbation of dynamical systems.} Springer NY (1984)

\bibitem{F1} Freidlin, M., I., {\it On stable oscillations and Equilibrium Induced by Small Noise.}, J. of Stat. Phys., Vol. 103, Nos. 1/2, 283-300 (2001)


\bibitem{F2} Freidlin, M., I., {\it On stochastic perturbation of dynamical systems with fast and slow components.}, Stochastics and Dynamics, Vol. 1, N. 2 , 261- 281 (2001)

\bibitem{G} Gammaitoni, L., Hanggi, P., Jung, P., Marchesoni, F., {\it Stochastic Resonance.} Rev. of Modern Physics 70 (1) 223-287 (1998)

\bibitem{HI} Herrmann, S., Imkeller, P., {\it The exit problem for diffusion with time periodic drift and stochastic resonance.}, Annals of Applied Probability 15 (1A) 39-68 (2001)

\bibitem{HM} Hitczenko, P.,  Medvedev , G.S.  {\it Bursting Oscillations induced by small noise.} Preprint arXiv:0712.4074v1 [nlin.AO] 25 Dec 2007


\bibitem{KS} Keener, Sneyd, {\it Mathematical Physiology. }  Mathematical Biology. Interdisciplinary Applied Mathematics, Springer (1998)

\bibitem{M} Murray, J., D., Mathematical Biology. Vol. 1 and 2, 2nd edition. Springer (2002)

\bibitem {N} Nagumo, J., S., Arimoto, S., Yoshizawa, S., {\it An active pulse transmission line simulating nerve axon.} Proc. IRE, 50, 2061-2071 (1962)

\bibitem{Pi} Piquet, C., Personnal communication.

\end{thebibliography}
\end{document}